\documentclass{moriond}

\def\be{\begin{equation}}
\def\ee{\end{equation}}
\def\bea{\begin{eqnarray}}
\def\eea{\end{eqnarray}}

\newcommand{\Photo}{}

\begin{document}
\vspace*{4cm}

\title{Electroweak baryogenesis in a scale invariant model and Higgs phenomenology
\footnote{Talk given at Rencontres de Moriond QCD and High Energy Interactions 2015, 
March 21-28, 2015, La Thuile, Italy.}}

\author{Kaori Fuyuto$^1$ and Eibun Senaha$^{1,2}$~\footnote{speaker}}
\address{$^1$Department of Physics, Nagoya University, Nagoya 464-8602, Japan, \\
$^2$Department of Physics and Center for Mathematics and Theoretical Physics, National Central University, Taoyuan, 32001, Taiwan}

\maketitle

\abstracts{
We study the electroweak phase transition and the critical bubble in the scale-invariant 
two Higgs doublet model taking the recent LHC data into account.
The sphaleron energy in this model is evaluated for the first time.
It is found that the strong first-order electroweak phase transition is the inevitable consequence
to be consistent with the observed 125 GeV Higgs boson.
In such a case, the signal strength of the Higgs decay to two gammas 
and the triple Higgs boson coupling could deviate from the SM values by $-10$\% 
and $+82$\%, respectively.
}

\section{Introduction}
Establishment of the Higgs sector is one of the primary issues in particle physics.
In 2012, a scalar boson was discovered at the Large Hadron Collider (LHC), 
and the mass of the particle has been determined with 0.2\% accuracy, 
$m_H=125.09\pm 0.21~({\rm stat.})\pm 0.11~({\rm syst.})$ GeV~\cite{Aad:2015zhl}.
Clarifying the properties of the particle is as important as its discovery 
since the discovered particle must have the important roles if it is really the Higgs boson, namely,
the origins of the mass generation and the electroweak symmetry (EW) breaking.
The experimental proof of the former is possible by measuring the Higgs boson couplings
to the gauge bosons and the fermions precisely, and the LHC experiment 
is now accessing those couplings. 
The latter can be clarified by reconstructing the Higgs potential.
In particular, the measurement of the triple Higgs boson coupling is enormously important 
since it can exist only after EW symmetry is broken.
So far, we know much less about the Higgs potential.

The EW symmetry can be broken if a tachyonic mass arises, which applies in the standard model (SM).
On the other hand, as pointed out by Coleman and Weinberg~\cite{Coleman:1973jx}, 
quantum corrections could also induce the EW symmetry breaking in massless theories.
One of the cosmological implications of such classical scale-invariant theories 
is that the EW phase transition (PT) is first order,
which is needed for successful EW baryogenesis (BG)~\cite{Kuzmin:1985mm}.
As explicitly demonsrated by Funakubo et al, the scale-invariant two Higgs doublet model (SI-2HDM) 
accommodates the strong first-order EWPT~\cite{Funakubo:1993jg}.
However, in their analysis the masses of the Higgs boson and the top quark were not fixed to
their observed values since those particles were not discovered at that time.

In this talk, we update the previous analysis in the light of the LHC data, 
and briefly discuss the phenomenological consequences in connection with Higgs physics.~\cite{Fuyuto:2015jha}

\section{Model}
The Higgs potential in the SI-2HDM is given by~\cite{Inoue:1979nn}
\bea
V_0
=\frac{\lambda_1}{2}(\Phi^{\dagger}_1\Phi_1)^2+\frac{\lambda_2}{2}(\Phi^{\dagger}_2\Phi_2)^2+\lambda_3(\Phi_1^{\dagger}\Phi_1)(\Phi_2^{\dagger}\Phi_2) 
+\lambda_4(\Phi^{\dagger}_1\Phi_2)(\Phi^{\dagger}_2\Phi_1) 
+\bigg[\frac{\lambda_5}{2}(\Phi^{\dagger}_1\Phi_2)^2+{\rm h.c.} \bigg],
\eea
where the mass terms are forbidden by the classical scale invariance, and
$Z_2$ symmetry is imposed to avoid the Higgs-mediated flavor-changing neutral current 
processes at tree level.
From the stationary conditions, one gets 
\bea
\tan^2\beta = \left(\frac{v_2}{v_1}\right)^2=\sqrt{\frac{\lambda_1}{\lambda_2}}, \quad\sqrt{\lambda_1\lambda_2}+\lambda_{345}=0,
\label{flat_cond}
\eea
where $v_1=v\cos\beta$ and $v_2=v\sin\beta$ with $v\simeq246$ GeV, 
and $\lambda_{345}=\lambda_3+\lambda_4+\lambda_5$.
We analyze the radiative EW symmetry breaking along the flat direction
using the Gildener-Weinberg method~\cite{Gildener:1976ih}.
The tree-level effective potential is
\bea
V_0(\varphi_1, \varphi_2)
= \frac{\lambda_1}{8}\varphi_1^4+\frac{\lambda_2}{8}\varphi_2^4
	+\frac{\lambda_{345}}{4}\varphi_1^2\varphi_2^2,
\eea
where $\varphi_{1,2}$ are the classical background fields.
Eq.~(\ref{flat_cond}) indicates that the energy of the minimum of $V_0$ is zero.
Furthermore, from Eq.~(\ref{flat_cond}), it follows that the determinant of the mass matrix of 
the CP-even Higgs bosons is zero.
Therefore, the tree-level potential has the flat direction.
The massless scalar is the consequence of the classical scale invariance.
The Higgs boson mass is generated after the EW symmetry is broken. Explicitly, one finds
\bea
m^2_h = \frac{1}{8\pi^2 v^4}
\Big[
	m_H^4 + m_A^4+2m_{H^\pm}^4 + 6m_W^4+3m_Z^4-12(m_t^4+m_b^4)
\Big].\label{Higgs_mass}
\eea
Note that $m_h^2$ becomes negative if the heavy Higgs bosons ($H$, $A$, $H^\pm$) are absent. 

%
%
\section{Sphaleron and Critical bubble}
In the EWBG mechanism, the baryon number ($B$) is created by expanding Higgs bubbles. 
$B$ can survive after the EWPT if the sphaleron process in the broken phase is quenched.
Here, as the sphaleron decoupling condition, we adopt 
$
\Gamma_B^{(b)}(T)\simeq ({\rm prefactor})e^{-E_{\rm sph}(T)/T} 
< H(T),
$
where $\Gamma_B^{(b)}$ denotes the sphaleron rate in the broken phase, which is
exponentially suppressed by $E_{\rm sph}(T)/T$, with $E_{\rm sph}(T)$ being the sphaleron energy
at a temperature $T$.
$H(T)$ is the Hubble parameter at $T$. 
We parametrize the sphaleron energy as $E_{\rm sph}(T)=4\pi v(T)\mathcal{E}(T)/g_2$, 
where $g_2$ denotes the SU(2) gauge coupling constant. 
The sphaleron decoupling condition then takes the form
\bea
\frac{v(T)}{T} > \frac{g_2}{4\pi \mathcal{E}(T)}
\Big[
	42.97+\mbox{log corrections}
\Big]\equiv \zeta_{\rm sph}(T).\label{sph_dec}
\eea
The log corrections mainly come from the fluctuation determinants about the sphaleron configuration,
which will be dropped in our numerical evaluation of $\zeta_{\rm sph}$ since they are subleading. 
We evaluate $v_C/T_C$ and $\zeta_{\rm sph}(T_C)$ numerically, where
$T_C$ is a temperature at which the effective potential has two degenerate vacua,
and $v_C$ is the Higgs vacuum expectation value at $T_C$.
(For a recent study on $\zeta_{\rm sph}(T_C)$ in the SM with a real singlet scalar, 
see Ref.~\cite{Fuyuto:2014yia}.)

If the supercooling is large, the use of the above criterion would not be appropriate.
As is done in the previous study, we also estimate a nucleation temperature ($T_N$), which is 
defined by
\begin{eqnarray}
\Gamma_N(T_N)/H^3(T_N)=H(T_N), \label{def_Tn}
\end{eqnarray}
where $\Gamma_N$ is the bubble nucleation rate per unit volume per unit time at $T_N$.
It should be emphasized that it is impossible to convert the entire region into the broken phase 
by only one bubble nucleated within the horizon volume.
Therefore, the nucleation temperature defined by Eq.~(\ref{def_Tn}) 
should be thought as an upper bound of the temperature at which the EWPT develops.

In studying Eqs.~(\ref{sph_dec}) and (\ref{def_Tn}), we use the following resummed
effective potential
\bea
V_{\rm eff}(\varphi,T)=\sum_in_i
\Bigg[ \frac{\bar{M}^4_i(\varphi,T)}{64\pi^2}\left(\log\frac{\bar{M}^2_i(\varphi,T)}{\bar{\mu}^2}-c_i\right) 
+\frac{T^4}{2\pi^2}I_{B,F}\left(\frac{\bar{M}^2_i(\varphi,T)}{T^2}\right)\Bigg],
\label{resummed_Veff}
\eea
with
$
I_{B,F}(a^2)=\int^{\infty}_0dx~x^2\log\left(1\mp e^{-\sqrt{x^2+a^2}}\right),
$
where $\bar{M}^2_i(\varphi,T)$ are the field-dependent boson masses with thermal corrections
~\cite{Fuyuto:2015jha}. 

\section{Results}
\begin{figure}[t]
\center
\includegraphics[width=7cm]{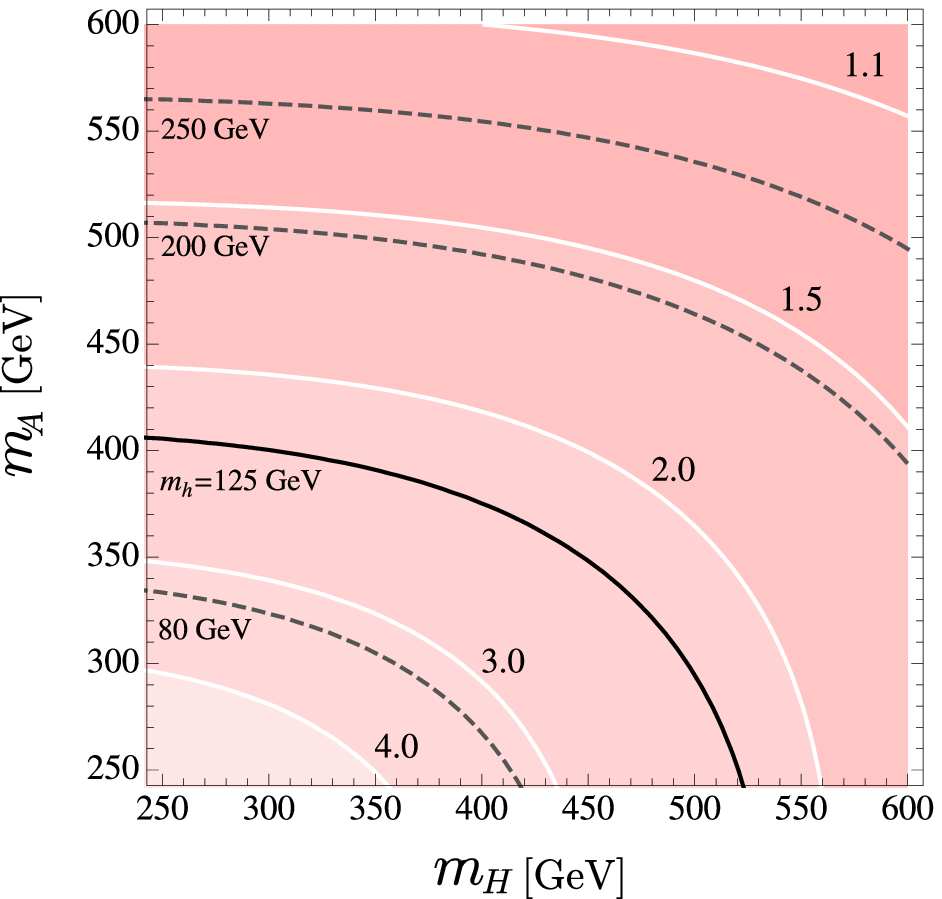}
\hspace{0.5cm}
\includegraphics[width=7cm]{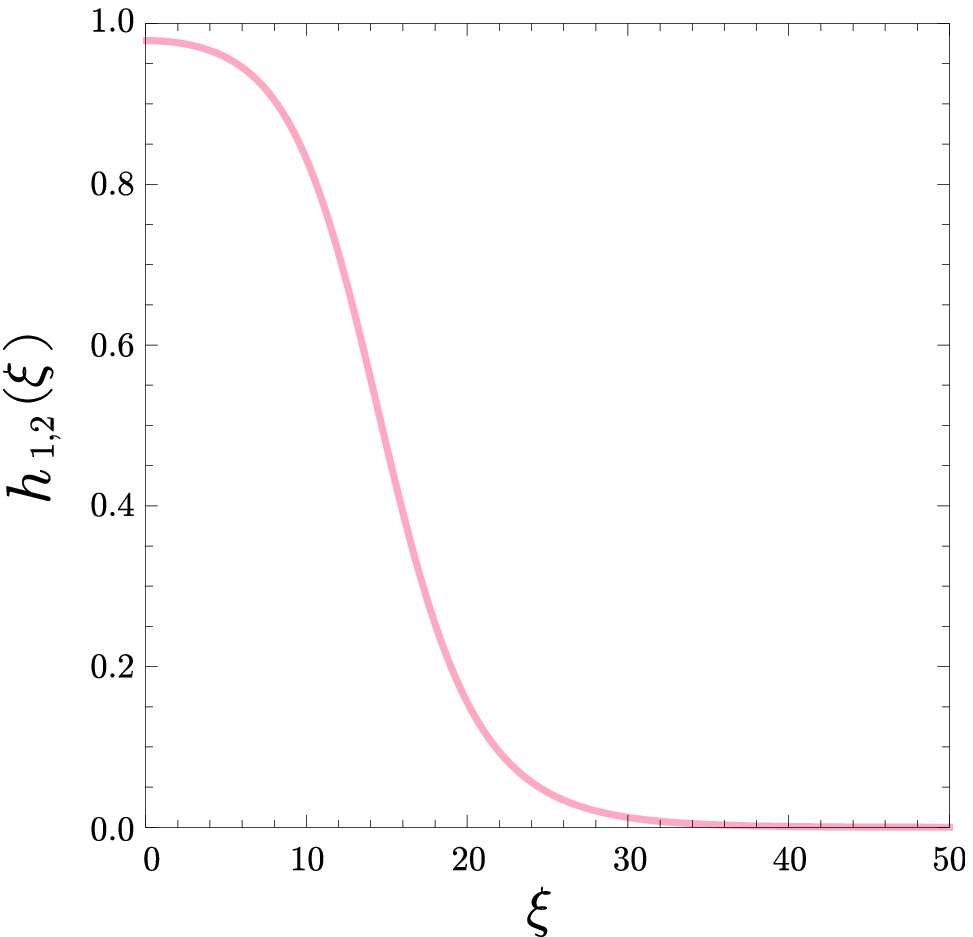} 
\caption{(Left panel) Contours of $m_h$ and $v_C/T_C$ in the $(m_H,~m_A)$ plane. 
The solid line in black represents $m_h=125$ GeV.
The each contour in white shows $v_C/T_C=1.1$, 1.5, 2.0, 3.0 and 4.0 from top to bottom.
(Right panel) The bubble wall profile at $T_N$.}
\label{fig:EWPT}
\end{figure}

In the left panel of Fig.~\ref{fig:EWPT}, $m_h$ and $v_C/T_C$ 
are plotted in the $(m_H,m_A)$ plane.
As seen from Eq.~(\ref{Higgs_mass}), $m_h$ goes up according 
as the heavy Higgs boson masses increase.
The black solid line corresponds to $m_h=125$ GeV.
In other words, the 125 GeV Higgs predicts the scale of the heavy Higgs bosons.
We overlay $v_C/T_C$ denoted by the white contours. From top to bottom, $v_C/T_C=1.1$,
1.5, 2.0, 3.0 and 4.0. It is concluded that the 125 GeV Higgs boson inevitably leads to
the strong first-order EWPT in the SI-2HDM.

In the right panel of Fig.~\ref{fig:EWPT}, the bubble wall profile at $T_N$ is shown. 
Unlike the minimal supersymmetric SM case~\cite{Funakubo:2009eg}, 
the bubble wall width is thinner in the SI-2HDM.

As a benchmark, we take $m_H=m_A=m_{H^{\pm}}=382$ GeV. 
Our findings are listed in Table~\ref{benchmark}.
The sphaleron decoupling condition is satisfied at $T_N$.
In this case, the signal strength of the Higgs boson decay to 2 gammas ($\mu_{\gamma\gamma}$) 
is reduced by 10\%
owing to the charged Higgs boson loop~\cite{Ginzburg:2002wt},
and the deviation of the $hhh$ coupling from the SM value ($\Delta\lambda_{hhh}$) 
is about +82\%.
The more detailed discussions on the phenomenology 
may be found in Refs.~\cite{Hill:2014mqa,Dorsch:2014qja}.

\begin{table}[t]
\center
\begin{tabular}{|c|c|}
\hline
$v_C/T_C$ &  211~GeV/91.5~GeV~=~2.31 \\
$\zeta_{\rm sph}(T_C)$ & 1.23  \\
\hline
$v_N/T_N$ &  229~GeV/77.8~GeV~=~2.94 \\
$\zeta_{\rm sph}(T_N)$ & 1.20  \\
$E_{\rm cb}(T_N)/T_N$ & 151.7 \\
\hline
$\kappa_V$ & 1.0\\
$\kappa_f$ & 1.0\\
$\mu_{\gamma\gamma}$ & 0.90 \\
$\Delta\lambda_{hhh}$ & 82.1$\%$ \\
\hline
$\Lambda$ & 6.3~TeV \\
\hline
\end{tabular}
\caption{The results in our benchmark scenario ($m_H=m_A=m_{H^{\pm}}=382$ GeV) are summarized . 
For the evaluation of the cutoff scale $\Lambda$, $\tan\beta=1$ is chosen as a reference value.}
\label{benchmark}
\end{table}

\section{Summary}
In this talk, the EWPT and the critical bubble in the SI-2HDM were revisited
in the light of the LHC data. 
We also estimated the sphaleron decoupling condition in this model for the first time.
To be consistent with 125 GeV Higgs boson, the EWPT is inevitably strongly first order.
Some of phenomenological consequences of this model are 
$\mu_{\gamma\gamma}=0.9$ and $\Delta\lambda_{hhh}=+82.1$\%.

\section*{Acknowledgments}

E.S. thanks the organizers for their kind invitation and great hospitality at the workshop.
E.S. is supported in part by the Ministry of Science and Technology of R. O. C. under Grant No. MOST 104-2811-M-008-011.
The work of K.F. is supported by Research Fellowships of the Japan Society
for the Promotion of Science for Young Scientists, No. 15J01079.


\section*{References}

\bibliography{senaha}


\end{document}